\documentclass[superscriptaddress,twocolumn,showpacs,preprintnumbers,amsmath,amssymb]{revtex4}

\usepackage{graphicx}
\usepackage{dcolumn}
\usepackage{bm}

\begin{document}

\title{The origin of fine structure in magnetization curve of $\alpha$-CoV$_2$O$_6$}

\author{Yu.B.Kudasov}
\email{yu_kudasov@yahoo.com}
\affiliation{Sarov Physics and Technology Institute, National Research Nuclear University ``MEPhI'',
Dukhov str. 6, Sarov, 607186, Russia}
\affiliation{National Research Nuclear University ``MEPhI'', Kashirskoe shosse 31, Moscow, 115409, Russia}
\affiliation{Russian Federal Nuclear Center - VNIIEF, Pr. Myra 37, Sarov, 607188, Russia}
\author{R.V.Kozabaranov}
\affiliation{Sarov Physics and Technology Institute, National Research Nuclear University ``MEPhI'',
Dukhov str. 6, Sarov, 607186, Russia}
\affiliation{National Research Nuclear University ``MEPhI'', Kashirskoe shosse 31, Moscow, 115409, Russia}

\date{\today}
\begin{abstract}
Multiple field-induced plateaus in $\alpha$-CoV$_2$O$_6$ at low temperatures were revealed earlier by M. Lenertz et al. [J. Phys. Chem. C 115, 17190 (2011)] and carefully investigated recently by M. Nandi and P. Mandal [J. Appl. 
Phys. 119, 133904 (2016)]. 
Four equidistant steps were observed in the magnetization curve. We present a model to describe this phenomenon. 
A magnetic structure
of this substance is formed by highly anisotropic triangular lattice of Ising chains running along the $\bf{b}$ axis.
Due to a three-fold degeneracy of
three-sublattice magnetic ordering, domain boundaries appear. Their transformation under magnetic field variation leads
to two additional steps in the 1/3 magnetization plateau and gives rise to complex magnetic behavior observed experimentally. 
The domain structure in $\alpha$-CoV$_2$O$_6$ occurs to be strongly anisotropic because a lifetime of the metastable states
depends greatly on the configuration orientation. A strong dependence of the magnetization curve on magnetic field sweep time is predicted.
\end{abstract}


\maketitle

Ising spin-chain compounds have drawn considerable attention due to unusual magnetic behavior \cite{kudasovUFN, mekata, drillon, hardy4}.
A complicated hierarchy of magnetic interactions determines main features of these systems. The strongest of them act along the chains. That is why in the framework of the rigid-chain model \cite{kudasovPRL} the
chains are considered as elements of the magnetic structure with two possible magnetization states. Weak antiferromagnetic (AFM)
interchain interactions in the triangular lattice of chains lead to the geometric frustration and complex phase diagram. At high temperatures
a partially disordered antiferromagnetic (PDA) phase appears \cite{mekata}. Very weak next-to-the-nearest interchain interactions or anisotropy 
partially lift the degeneracy of the ground state and produce various low-temperature magnetic structures: the three-sublattice (3SL) one in CsCoCl$_3$ and related compounds, stripes
in Sr$_5$Rh$_4$O$_{12}$ \cite{cao,kudasovJETPL}, and strongly disordered Wannier's state \cite{wannier} in Ca$_3$Co$_2$O$_6$ \cite{drillon, hardy1, kudasovPRB}.  

One of the most striking features observed in the spin-chain compounds is a multi-step magnetization curve in Ca$_3$Co$_2$O$_6$ \cite{drillon,hardy1} and Sr$_3$HoCrO$_6$ \cite{hardy4}. 
The plateau 1/3 at high temperatures is related to transition to the 3SL phase. At low temperatures two additional steps appear at the plateau. They are assumed to be a consequence
of a three-fold degeneracy of the 3SL configuration which gives rise to a domain structure \cite{kudasovUFN,kudasovPRB2}. This is a long-lived metastable state which determines a slow magnetic dynamics in Ca$_3$Co$_2$O$_6$ 
\cite{kudasovUFN,hardy1}.  

\begin{figure}
\includegraphics[width=0.45\textwidth]{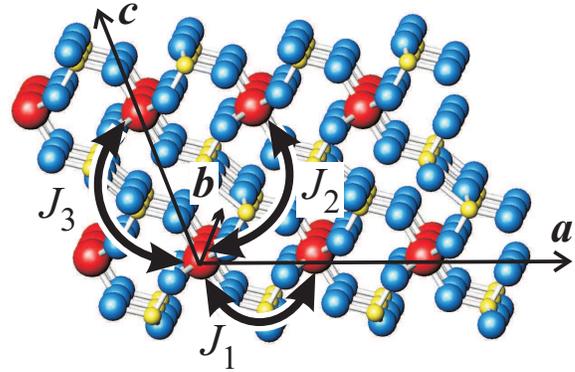}
\caption{\label{f1} (color online) The crystal structure of $\alpha$-CoV$_2$O$_6$. The large red, smaller blue, and the smallest yellow spheres denote the Co$^{2+}$, O$^{2-}$, and V$^{5+}$ ions correspondingly. The arrows show 
interchain interactions.}
\end{figure}

In this Brief Report we discuss the multi-step behavior in CoV$_2$O$_6$ revealed in Ref.~\cite{lenertz} and thoroughly investigated recently \cite{nandi}.
This compound crystallizes in the two allotropic forms  \cite{lenertz}: the branneritelike monoclinic structure with the space group C2/m which
is schematically shown in Fig.~\ref{f1} ($\alpha$-phase) and the triclinic structure
with the P-1 space group ($\gamma$-phase). They have rather similar magnetic behavior, however in the present article we discuss the $\alpha$-CoV$_2$O$_6$ only.
Magnetic Co$^{2+}$ ions in the $\alpha$ phase are placed in the well separated chains 
made up of edge-sharing distorted CoO$_6$ octahedra. The chains run along the $\bf{b}$ axis. Pentavalent vanadium ions are nonmagnetic and located in edge-shared VO$_5$ square-pyramids between the chains being involved in 
interchain magnetic interactions.

Magnetic measurements of $\alpha$-CoV$_2$O$_6$ revealed
an Ising-type anisotropy with the easy axis along $\bf{c}$ axis \cite{he}. It arises from cooperative effects of the crystal electric field of the distorted octahedra and spin-orbit coupling \cite{bkim}. This combination also 
leads to unusually high orbital moment of Co$^{2+}$ ions. The magnetic structure is formed by the strong ferromagentic (FM) in-chain coupling between
the Co$^{2+}$ ions, and weak AFM interchain interactions. They lead to the AFM state below $T_N=15$~K.
The lattice of the chains can be considered as a distorted 2D triangular lattice with three AFM coupling
constants between the nearest neighboring chains $J_1$, $J_2$, and $J_3$ which act along
$[100]$, $[101]$, and $[001]$ directions, respectively (see Fig.~\ref{f1}). 
The values of $J_1$, $J_3$ although are different but comparable, $J_2$ is much weaker than $J_1$
and $J_3$. According to Ref.~\cite{lenertz2} the estimations of the coupling constant are $J_1=0.356$~meV, $J_2=0.051$~meV, and $J_3=0.463$~meV.
As was shown in Ref.~\cite{saul} interactions between the next to the nearest neighbor are rather weak and can be omitted. 

Magnetization curves of $\alpha$-CoV$_2$O$_6$ below $T_N$ demonstrated two steps separating plateaus with zero magnetization ($M=0$) at low magnetic field, one third of the saturation magnetization ($M=1/3$), and the saturated 
state ($M=1$). This structure was observed from 12~K to 5~K. The neutron scattering \cite{markkula,lenertz2} and Monte Carlo simulation showed that the first plateau 
corresponds to
the stripe magnetic structure, and the 1/3 plateau does to three sublattices (3SL) configurations. 
The Monte Carlo simulation by the Metropolis algorithm in some cases gave additional features
in the 1/3 plateau. However, it should be mentioned that this technique
fails to relax correctly a trial state into the equilibrium state on the AFM triangular Ising model and
the Wang-Landau algorithm should be applied instead \cite{qin}. The Wang-Landau simulation produces
the flat 1/3 plateau in case of $\alpha$-CoV$_2$O$_6$ \cite{yao1}.

\begin{figure}
\includegraphics[width=0.2\textwidth]{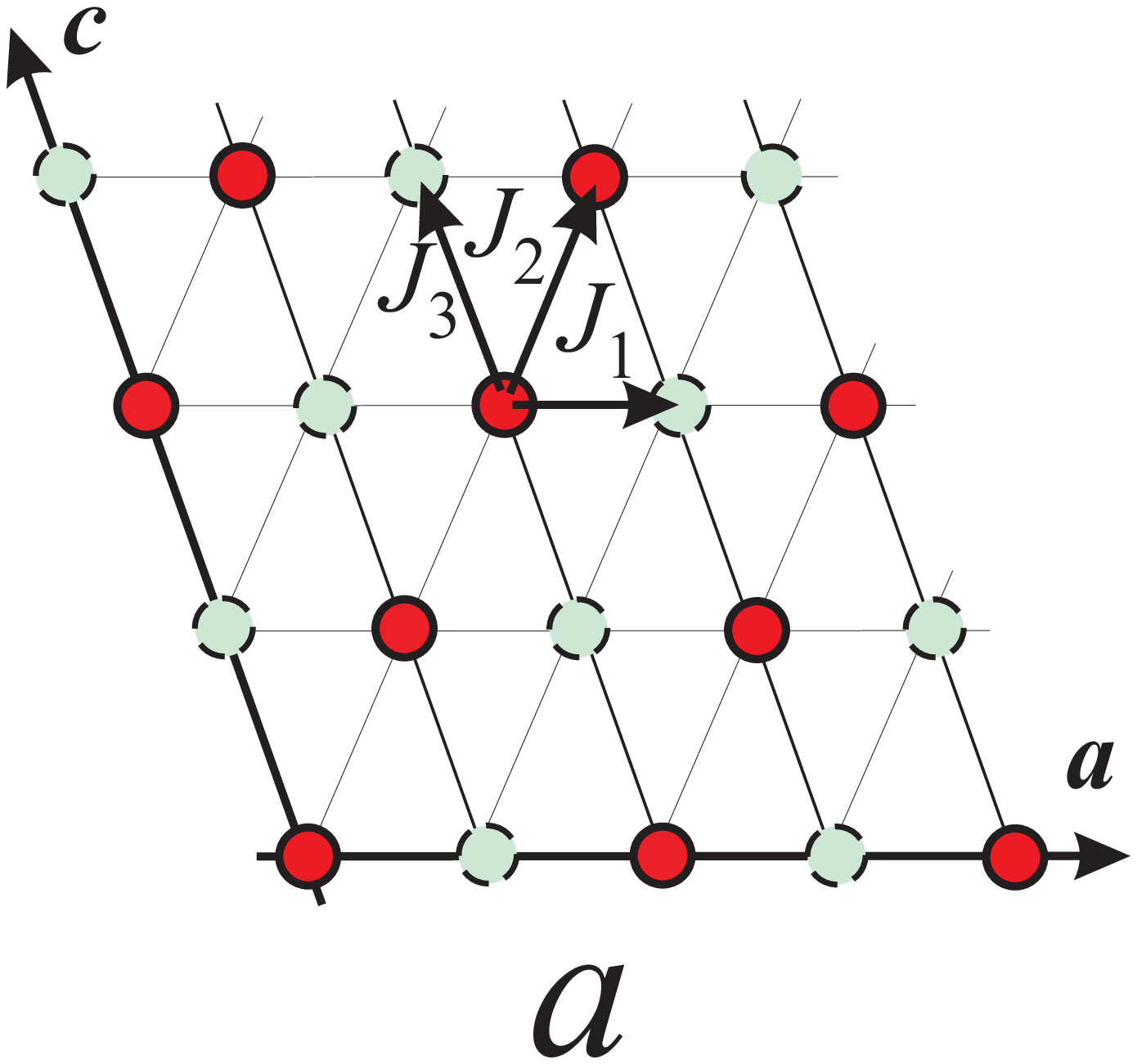}
\includegraphics[width=0.2\textwidth]{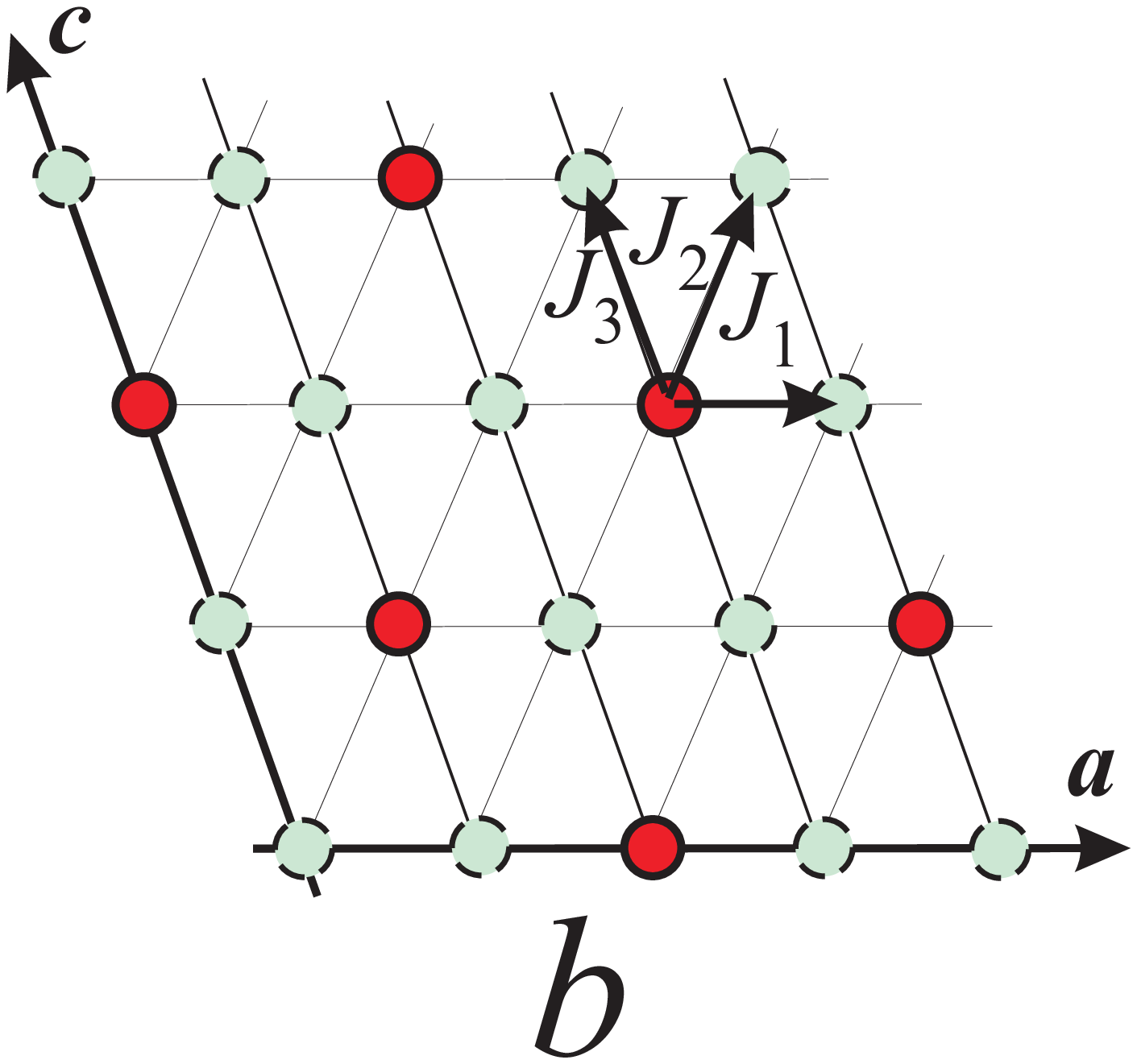}
\caption{\label{f2} (color online) Magnetic structures of stripe (a) and 3SL phases. The solid (red) and dash (blue) circles denote spin-down and spin-up (along the magnetic field) chain states, repectively.}
\end{figure}
 
Two additional steps steps in the 1/3 plateau were revealed at very low temperatures below 5~K \cite{lenertz}. The
four magnetization steps are equidistant on the magnetic field (1.6, 2.2, 2.8, and 3.4~T) \cite{lenertz,nandi} similarly to Ca$_3$Co$_2$O$_6$ despite the fact the 2D lattice of chains is strongly anisotropic. This phenomenon is 
worthy of thorough investigation.

The rigid-chain model for $\alpha$-CoV$_2$O$_6$ was formulated by Yao as following \cite{yao1}     
\begin{eqnarray}
\hat{H}=\sum_{<ij>}{J_{ij} \sigma_i \sigma_j} - h M  \label{Ham}
\end{eqnarray}
where $\sigma_i = \pm 1$ is the $c$-axis projection of the $i$-th chain magnetic moment, $M=\sum_{i}{\sigma_i}$ is the total magnetization,  $<\ldots>$ denotes the summation over the nearest neighboring chains only,
$J_{ij}$ is a positive constant which assumes values $J_1$, $J_2$, or $J_3$ ($J_1 \approx J_3$, $J_2 < J_1$, 
$J_2 < J_3$) depending on the bond orientation,
$h$ is the magnetic field. 

At the zero magnetic field the equilibrium state is the stripe structure oriented along the $[101]$ direction as shown in the Fig.~\ref{f2}a. While magnetic field increasing a first order transition to the 3SL phase occur at the 
first
critical magnetic field \cite{yao1}

\begin{equation}
h_{c1}=2(J_1 +J_3) - 4J_2.
\label{hc1}
\end{equation}
The 3SL structure is shown in Fig.~\ref{f2}b.
Further increase of the magnetic field leads to the second transition to the saturated FM state at the second critical magnetic field \cite{yao1}

\begin{equation}
h_{c2}=2(J_1 +J_3 + J_2).
\label{hc2}
\end{equation}

\begin{figure}
\includegraphics[width=0.4\textwidth]{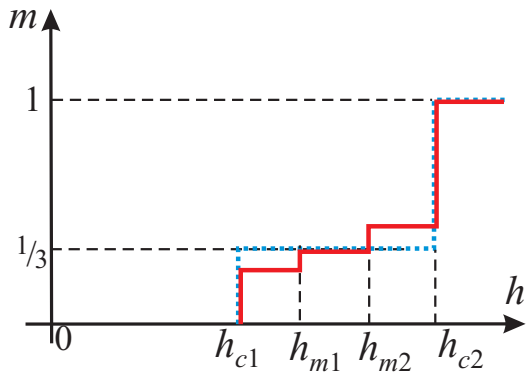}
\caption{\label{f3} (color online) A schematic view of the magnetization curves of $\alpha$-CoV$_2$O$_6$ in the increasing magnetic field: the dotted (blue) and the solid (red) lines correspond to the equilibrium and metastable 
regimes.}
\end{figure}

These two critical magnetic fields form the two steps in the equilibrium magnetization curve (Fig.~\ref{f3})
which was observed in the temperature range between 5 and 15~K. 

To discuss the multi-step behavior at low temperatures
one should take into account frozen metastable states. Necessary conditions of the metastability at the zero temperature have the following form \cite{kim,kudasovPRL}  
\begin{equation}
\sigma_i h_i\leq 0
\label{meta}
\end{equation}
where $h_i = \sum_{j(i)}{J_{ij}\sigma_j}-h$ is  the effective field for the $i$-th chain. This condition should be satisfied for all the chains.

At finite temperatures the lifetime $\tau$ of a metastable state can be estimated by the Glauber theory \cite{glauber} extended to 2D and 3D lattices \cite{buendia,kudasovUFN}. 
Following to it
one assumes that the chains interact not only with the nearest neighbors and external
magnetic field but also with a heat reservoir. It was shown that a hard type of the Glauber dynamics describes well the slow dynamics in frustrated spin-chain system \cite{kudasovUFN,kudasovPRB2,buendia}. 
In this case the probability of a spin flip of the $i$-th chain per time unit can be written down as
\begin{equation}
W_i=\frac{\alpha}{2} 
 \left[1-\sigma_i \cdot \tanh \left( - \frac{h_i}{k_B T}\right) \right]  
\label{Glauber}
\end{equation}
where $\alpha$ is the constant of the interaction of a chain with the heat reservoir, $k_B$ is the Boltzmann constant, 
$T$ is the temperature. Then the lifetime of the $i$-th chain state is $\tau_i=W_i^{-1}$. If $\tau \sim \alpha^{-1}$ the state is unstable, and $\tau >> \alpha^{-1}$ corresponds to a metastable (or stable) state.

\begin{figure}
\includegraphics[width=0.4\textwidth]{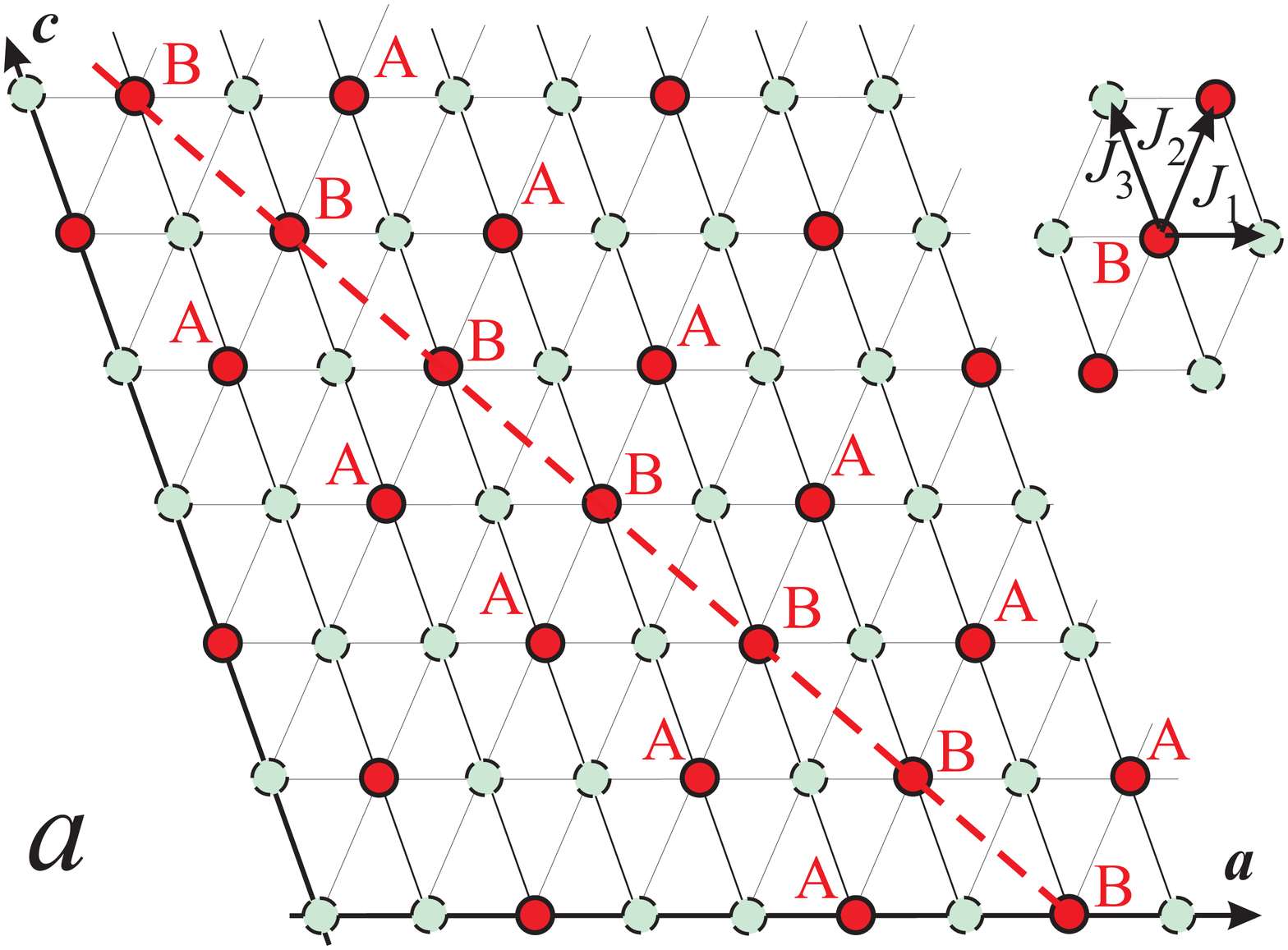}
\includegraphics[width=0.4\textwidth]{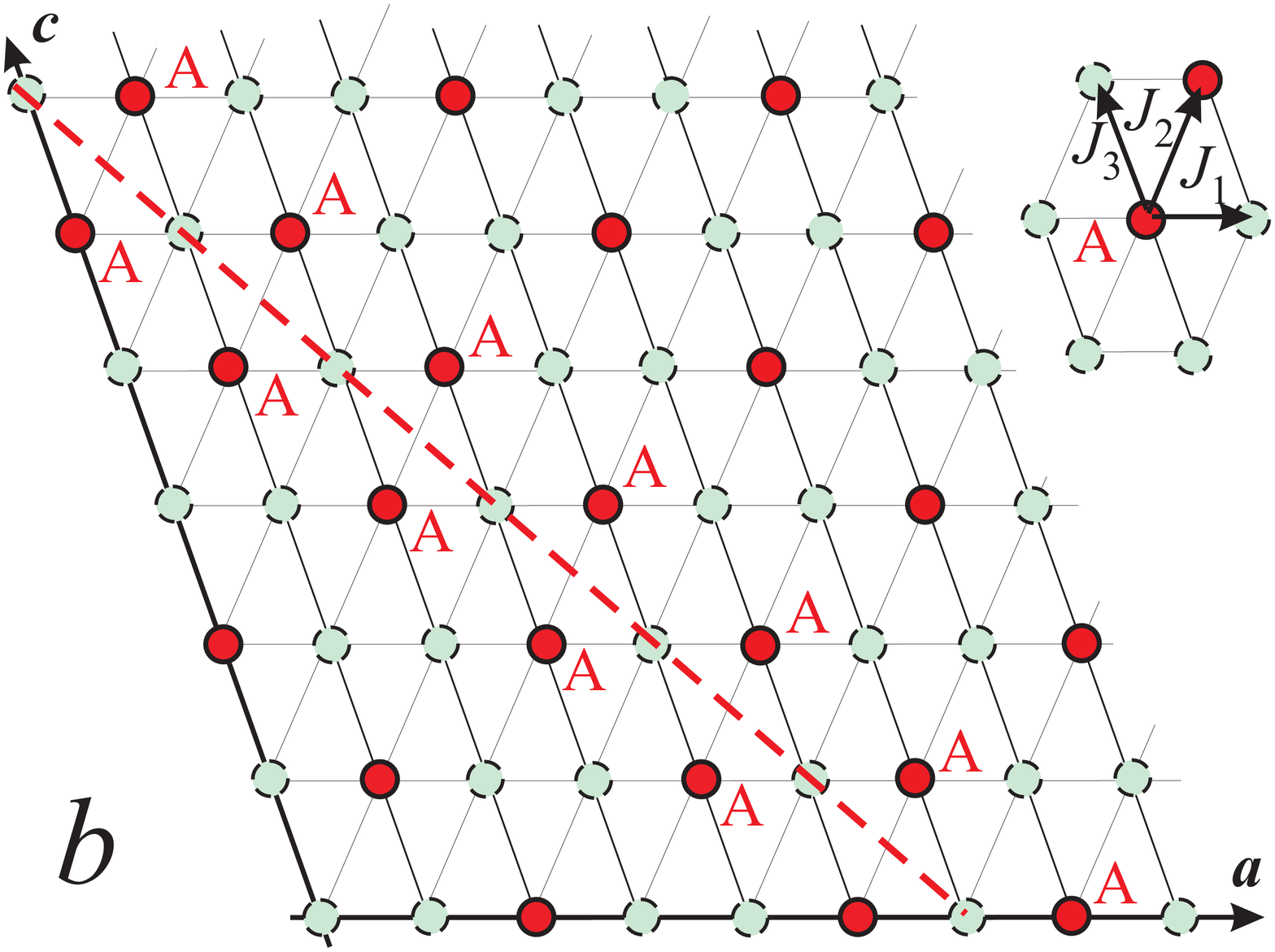}
\caption{\label{f4} (color online) Two possible structures of the domain boundaries in the 3SL phase in the (101) planes. A and B denote positions of spin-down chains with two different types of the nearest-neighbor 
environment.}
\end{figure}

There are two basic causes for the metastability just above $h_{c1}$. Firstly, using the condition Eq.~(\ref{meta})
it is easy to show that the stripe phase still exist as a metastable state up to the magnetic field
\begin{equation}
h_{m1}=2(J_1 +J_3) - 2J_2.
\label{hm1}
\end{equation}
The 3SL structure is three-fold degenerate, and when it starts growing from few nucleation centers domain boundaries
appear. Let us consider the domain boundaries lying in the (101) planes. The possible configurations are shown
in Fig.~\ref{f4}. Since the FM bonds for the spin-down chains in domain boundaries (A and B positions) are
along the $[101]$ direction with the weakest AFM coupling constant $J_2$ this orientation makes the most stable
domain boundaries. We discuss this point below. The configurations in Fig.~\ref{f4}a give an excess of spin-down chains as compared with monodomain 3SL structure. That is why just above $h_{c1}$ the magnetization of the 
metastable state is less than that of the equilibrium one. The chains in the position B become unstable at the magnetic
field determined by Eq.~(\ref{hm1}) that causes the additional step in the magnetization curve (see Fig.~\ref{f3}).  

The spin-down chains in the position A in Fig.~\ref{f4}b remain stable up to another critical field
\begin{equation}
h_{m2}=2(J_1 +J_3).
\label{hm2}
\end{equation}
At this point half of the spin-down chains in the A-A pairs (Fig.~\ref{f3}) flips to the spin-up state. 
This gives rise to an excess of spin-up chains. Therefore, there appears a magnetization step at $h_{m2}$
and the plateau with the magnetization greater than the equilibrium one. In the range between $h_{m2}$
and $h_{c2}$ spin-down chains are stable only in case of all the six nearest neighbors are in the spin-up
state. 

\begin{figure}
\includegraphics[width=0.25\textwidth]{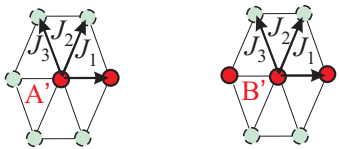}
\caption{\label{f5} (color online) Examples of possible short-lived metastable configurations.}
\end{figure}

Until now we discussed the domain boundaries lying in the $(101)$ plane. Let us consider the ones in other planes, e.g.
$(100)$, and more generally other types of metastable states. Then we investigate the stability of the chain in the
down state similar to the B position but with two nearest neighbors in the down positions along the $[100]$ direction
(B$'$ position see Fig.~\ref{f5}). The corresponding critical magnetic fields can be calculated by a coupling
constants permutation. Then the equation Eq.~(\ref{hm1})
is replaced by $h^{\prime} _{m1}=2(J_2 +J_3) - 2 J_1$. If $h^{\prime}_{m1}<h_{c1}$ the B$'$ states are unstable,
that is they exist only if $J_2 > 2/3 J_1$.

Let us compare the lifetime for different metastable states. The additional steps at the 1/3 plateau are observed at very low temperatures only when $k_B T$ is much less than any of $J_1$, $J_2$, and $J_3$. 
Then one can estimate the lifetime from Eq.~(\ref{Glauber}) as $\tau_i \approx \alpha^{-1} \exp{(|h_i| / k_B T )}$. Therefore  the lifetime is large as compared to $\alpha^{-1}$
in all the area of the metastablity except a narrow region in the vicinity of the critical field where $|h_i|$ is close to or less than $k_B T$. Applying this expression we obtain that the lifetime of a chain in the B position
larger than that in the B$'$ one by a factor of $\exp\left[2(J_1-J_2)/(k_B T)\right]>>1$. 

After a similar reasoning one can show that if a chain in the down state has one or two nearest neighbors in the
down states then any configuration, which differs from A or B positions (e.g. A$'$ or B$'$ in Fig.~\ref{f5}),
has exponentially small lifetime and therefore can be neglected.

It is interesting to mention that the four-step structure in the magnetization curve can appear in the two cases only: perfect triangular lattice (Ca$_3$Co$_2$O$_6$) and strongly anisotropic
triangular lattice ($\alpha$-CoV$_2$O$_6$). If the anisotropy is weak the critical fields for different orientation of domain boundaries slightly differ from one another.
This should smear the additional steps at the plateau 1/3.

It is possible to get a new relation between the coupling constants from the model developed above:
\begin{equation}
\frac{h_{c2}-h_{c1}}{h_{m2}}=\frac{3 J_2}{J_1+J3}.
\label{r}
\end{equation}
From here one can see that the value of $J_2$ obtained in Ref.~\cite{lenertz2} and mentioned above is
underestimated by a factor of 3.4 as compared to $J_1$ and $J_3$.

In conclusion, 
there are very few magnetic systems which demonstrate the four-step magnetization curves (e.g. Ca$_3$Co$_2$O$_6$ \cite{drillon, hardy1} and Sr$_3$HoCrO$_6$ \cite{hardy4}). The $\alpha$ phase of CoV$_2$O$_6$ is the only
one among them with the first step of the four-step structure shifted from the zero to finite magnetic field. The nature of
the additional steps in the plateau 1/3 observed in this compound stems from metastable states induced by domain boundaries in the 3SL phase. In contrast to Ca$_3$Co$_2$O$_6$ the lattice of chains in $\alpha$-CoV$_2$O$_6$
is strongly anisotropic. This leads to various possible metastable configurations with particular critical fields. However only few of them
with the special orientation
have the large lifetime which is sufficient for observation. It is exponentially small for the other configurations 
and therefore they can be neglected. That is why the domain structure in the 3SL phase should be
highly anisotropic, i.e. the domain boundaries of the types shown in Fig.~\ref{f4} lie in the (101) plane only. 
It is important to investigate the dependence of
the magnetization curve at low temperatures on the magnetic field sweep rate in detail similarly to Ca$_3$Co$_2$O$_6$. A simulation of the magnetization dynamics in the framework of Glauber's theory would be also useful for 
deep insight into the magnetic phenomena in $\alpha$-CoV$_2$O$_6$.  
We assume that at very slow magnetic field variation the magnetization curve should tend to
the two-step structure with the flat plateau 1/3.
In the opposite case of large sweep rate the four-step magnetization curve
also could be broken by the short-lived metastable states.


\end{document}